\documentclass[letter]{jpsj3}

\title{Mobility Analysis of FeTe Thin Films}

\author{
Ichiro TSUKADA$^{1,3}$\thanks{E-mail address: ichiro@criepi.denken.or.jp}, 
Masafumi HANAWA$^{1,3}$, 
Seiki KOMIYA$^{1,3}$, 
Ataru ICHINOSE$^{1,3}$, 
Takanori AKIIKE$^{2,3}$, 
Yoshinori IMAI$^{2,3}$, 
and Atsutaka MAEDA$^{2,3}$
}

\inst{
$^{1}$Central Research Institute of Electric Power Industry, 2-6-1 Nagasaka, Yokosuka, Kanagawa 240-0196, Japan \\
$^{2}$Department of Basic Science, The University of Tokyo, 3-8-1 Komaba, Meguro, Tokyo 153-8902, Japan \\
$^{3}$JST, TRIP, Sanbancho, Chiyoda, Tokyo 102-0075, Japan
}

\abst{
The Hall effect is investigated in detail for nonsuperconducting and 
superconducting FeTe thin films. 
The Hall coefficient commonly exhibits a sign reversal from positive 
in a high-temperature paramagnetic state 
to negative in a low-temperature antiferromagnetic state. 
Phenomenological analysis by a simple two-band Drude model indicates 
that hole mobility is significantly suppressed in the antiferromagnetic state. 
When suppression of the hole mobility is insufficient, 
superconductivity shows up in FeTe. 
This result strongly suggests that the itinerancy in both hole and electron 
channels is the essential factor for the occurrence of superconductivity 
in iron chalcogenide superconductors.
}

\kword{iron chalcogenide, FeTe, Hall effect, thin films, two-band model}

\begin{document}
\maketitle

%Introduction

Iron-pnictide 
\cite{Kamihara1} 
and iron-chalcogenide 
\cite{Hsu1} 
superconductors are interesting materials when compared with 
cuprate superconductors. 
In both compounds, superconductivity is induced by chemical substitution 
to a parent antiferromagnet suggesting a possible common mechanism 
resulting in a similar phase diagram. 
However, the nature of the antiferromagnetic (AFM) state is 
qualitatively different. 
Cuprates are characterized as single-band metals and become an insulator, 
while iron-based superconductors are characterized as multi-band metals,
\cite{Singh1,Mazin1,Kuroki1} 
and exhibit metallic behavior even in the AFM state.
\cite{Rotter1} 
Actually undoped parent compounds of iron-based superconductors are compensated 
metals, where both electrons and holes contribute to the electrical transport. 
Therefore, this difference should be carefully considered when discussing 
the phase diagram, and it is important to understand how electrons and holes 
contribute to the electrical transport and superconductivity.

The multi-band nature of iron-based superconductors has been predicted 
by several band calculations carried out for 
LaFeAsO,
\cite{Singh1} 
BaFe$_2$As$_2$,
\cite{Singh2} 
and Fe(Se,Te).
\cite{Subedi1} 
In LaFeAsO and BaFe$_2$As$_2$, superconductivity is induced by chemical 
substitution of elements with different valence number, 
and a rigid-band picture properly works.
\cite{Kohama1,Rullier1,Fang1} 
Isovalent Co substitution of Fe in BaFe$_2$As$_2$ is also considered to shift 
the Fermi level, and thus works as an electron doping to BaFe$_2$As$_2$.
\cite{Katayama1} 
In contrast to them, the role of chemical substitution in iron chalcogenides 
has remained unclear. 
Se and Te take the same valence state, 
and the band structures are quite similar to each other.
\cite{Subedi1} 
Hall effect measurements are one of the powerful techniques to trace 
how the electronic state changes from FeTe with Se substitution.
\cite{Tsukada1} 
In this letter, we report on the detailed Hall-effect measurements in the 
parent compound FeTe thin films, and try to apply two-band Drude picture 
to demonstrate an interplay of electrons and holes using resistivity 
and Hall coefficients data. 
Although the band calculation predicts the sufficient amount of electron and 
hole density in FeTe, difference of the magnitude of mobility is remarkable 
not only in the paramagnetic (PM) but also in the AFM states, 
which suggests the absence of interplay between two types of carriers.

%(Experimental)s

Thin film samples were prepared by pulsed-laser deposition using 
a stoichiometric FeTe sintered target.
\cite{Imai1,Tsukada1} 
We choose MgO(100) as a substrate following the results of substrate 
selection in the growth of Fe(Se$_{0.5}$Te$_{0.5}$) thin films.
\cite{Imai2} 
A metal mask was used to make the film in a six-terminal shape. 
Three FeTe films with different thicknesses are compared in the present study: 
The thickness of samples A, B, and C is 40, 400, and 165~nm, respectively. 
The crystal structure is characterized by an x-ray diffraction and transmission 
electron microscopy (TEM), and the charge transport is characterized 
by Physical Properties Measurement System (Quantum Design).

%(Experimental Results) 

%(XRD and RT1)

\begin{figure}
\begin{center}
\includegraphics*[width=120mm]{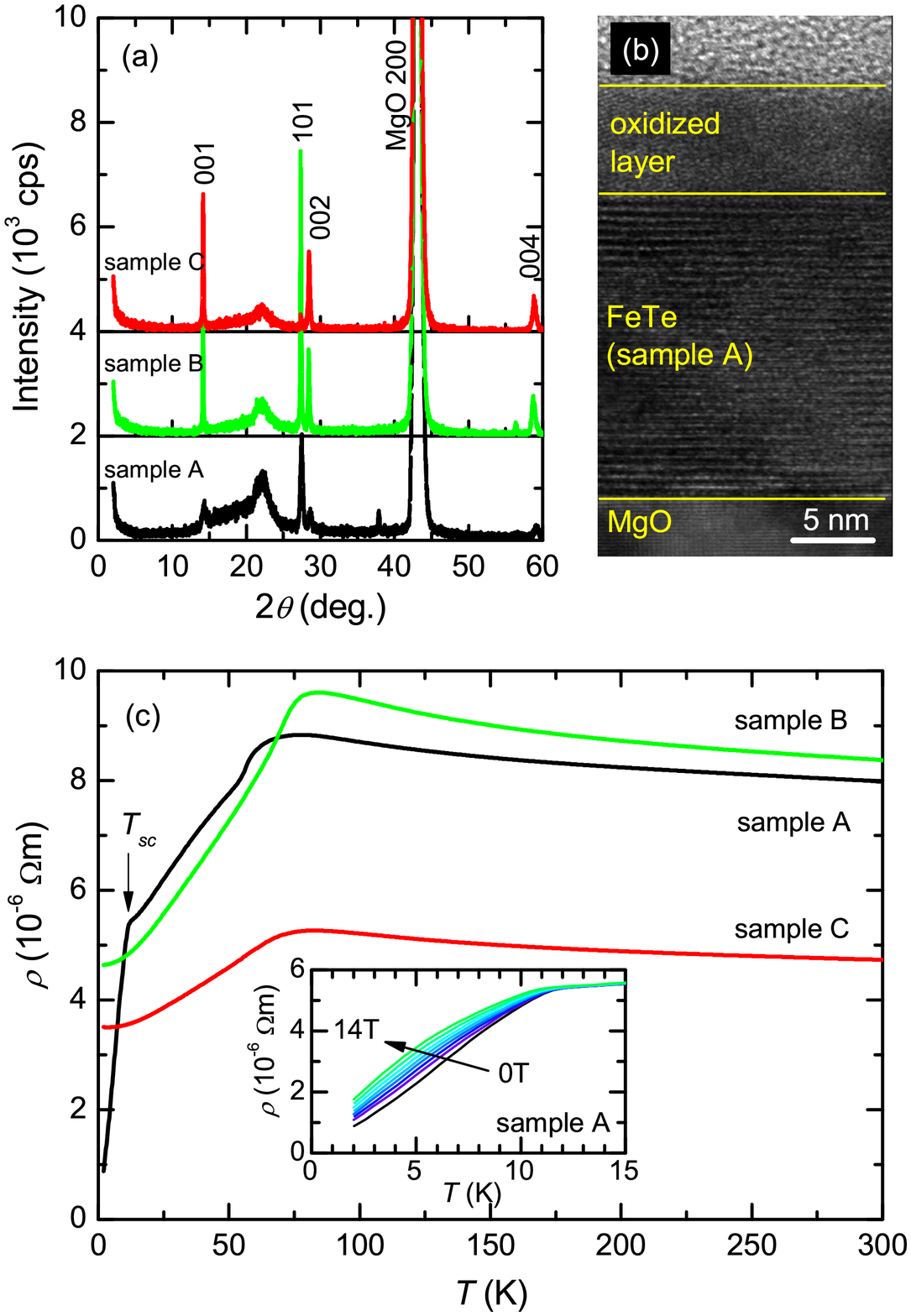}
\end{center}
\caption{(Color online) 
(a) X-ray diffraction of the three films. (b) Cross section TEM 
image of sample A. (c) Temperature dependence of resistivity 
of the three films. The inset shows magnetic field dependence 
of resistivity of sample A.
}
\label{Fig.1}
\end{figure}

Figure~\ref{Fig.1}(a) shows x-ray diffractions of the three samples. 
Sample C shows almost perfect $c$-axis orientation, 
while the other two contain the (101)-oriented domains. 
In sample A, the 101 reflection is far intense than the 00$l$ reflections. 
In sample B the intensity of the 00$l$ reflections becomes strong but 
still the portion of the (101)-oriented domains are not small. 
We should note, however, that even in sample A high-resolution transmission 
microscopy (TEM) observation demonstrates a highly $c$-axis oriented structure 
[Fig.~\ref{Fig.1}(b)]. 
We can see a sharp interface between MgO and FeTe similar to the case 
of Fe(Se$_{0.5}$Te$_{0.5}$) films on MgO,
\cite{Imai2} 
but simultaneously some deteriorated layer is observed at the surface 
probably due to oxidation during long-time expose in the air. 
The calculated $c$-axis lengths from the 00$l$ reflections are 
6.258~{\AA}, 6.300~{\AA}, and 6.285~{\AA} for samples A, B, and C, 
respectively.

Samples A and B exhibit higher resistivities than sample C does 
as shown in Fig.~\ref{Fig.1}(c), probably because of the admixture of the 
(101)-oriented domains. 
The difference is consistent with that $\rho$ along the FeTe layer is slightly 
lower than that perpendicular to it. 
It should be noted that $\rho$ of sample C almost equals to that of bulk single 
crystals ever reported,
\cite{Noji1} 
indicating that sample C exhibits an intrinsic in-plane electric transport. 
Trace of the tetragonal-to-monoclinic structural phase transition 
is not observed in samples B and C. 
In bulk samples, resistivity exhibits a discontinuous jump at $T_{tm}$ = 68~K,
\cite{Mizuguchi1,Noji1} 
but samples B and C exhibit only a rounded peak in $\rho$ around $T$ = 80~K 
followed by a rather smooth decrease to low temperatures 
as reported.
\cite{Mele1,Han1} 
However, except for the absence of sharp drop at $T_{tm}$, overall temperature 
dependence of $\rho$ is similar between the films and bulk crystals. 
Thus, crossover from tetragonal to monoclinic phases is likely to take place, 
and hence, we safely assume that an AFM long-range order is present 
in samples B and C at low temperatures. 
In contrast to them, $\rho$ shows a steep drop at $T$ = 57~K in sample A, 
indicating the presence of a first-order transition.

Another important feature is that sample A shows a steep drop of resistivity 
below 11~K. 
This can be attributed to the onset of superconductivity, because a typical 
suppression of the transition temperature ($T_c$) with external magnetic field 
is observed as shown in the inset of Fig~\ref{Fig.1}(d). 
The superconductivity was less obvious immediately after the growth of films, 
but showed up after two months, and then became unchanged. 
Thus, this superconductivity seems to be due to an aging effect 
discussed for Fe(Te,Se) polycrystalline samples.
\cite{Mizuguchi2,Mizuguchi3,Nie1} 
However, we cannot exclude a possible epitaxial strain effect 
discussed by Han {\it et al.,}
\cite{Han1} 
because the $c$-axis length is far shorter in sample A than in samples B and C. 
Whatever the reason is, our interest is focused on how the appearance 
of superconductivity is related to the normal-state electric transport.

%(Hall1)

\begin{figure}
\begin{center}
\includegraphics*[width=120mm]{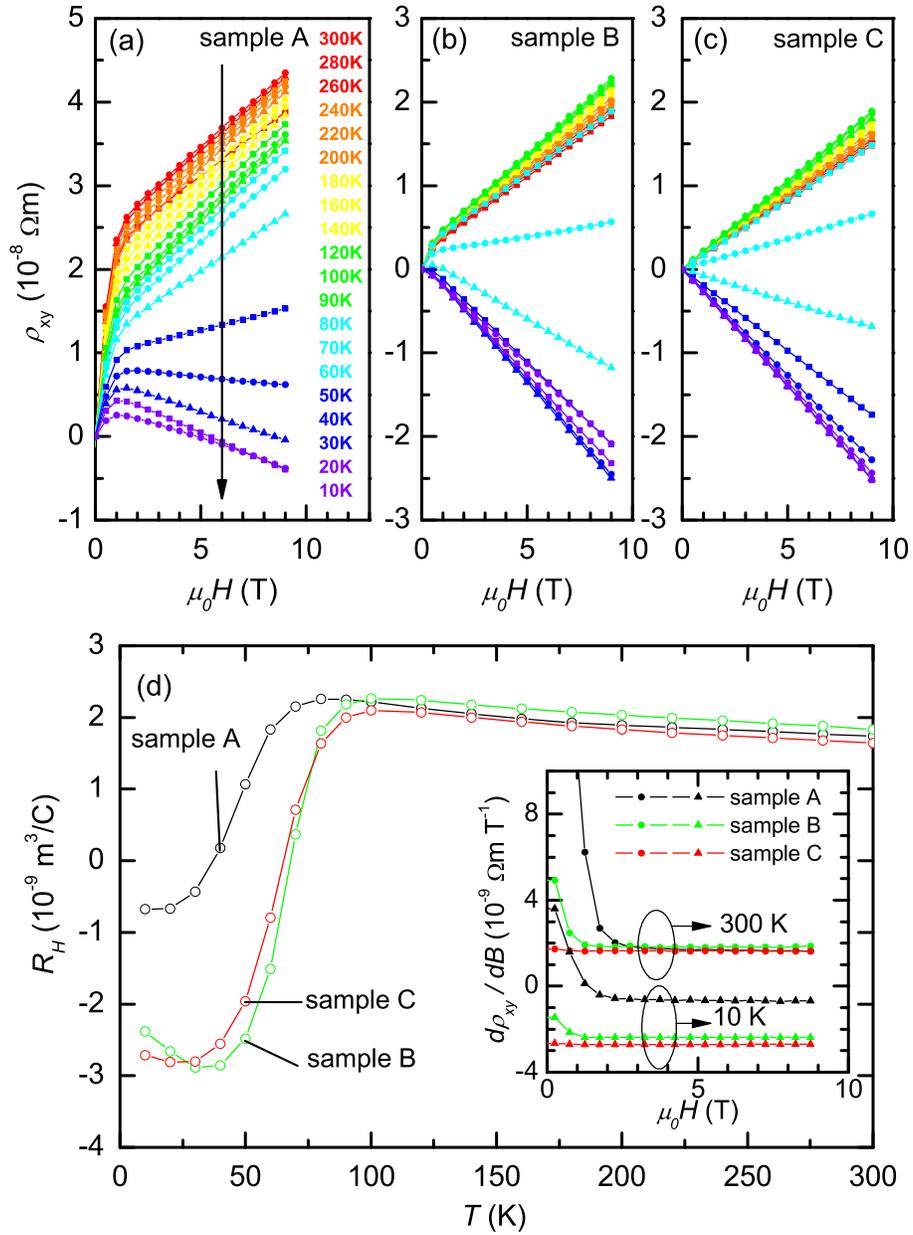}
\end{center}
\caption{(Color online) 
(a)-(c) Field dependence of $\rho_{xy}$ at different temperatures 
for the three films. 
Finite contribution of the anomalous Hall effect is detected in $\rho_{xy}$ 
of samples A and B, while $\rho_{xy}$ is almost linear to $B$ in sample C. 
(d) Temperature dependence of $R_H$ for the three films. 
The inset shows $d\rho_{xy} / dB$ at $T$ = 300 and 10~K.}
\label{Fig.2}
\end{figure}

The Hall effect is investigated in the same way as was performed 
to Fe(Se,Te) thin films.
\cite{Tsukada1} 
We first measure the transverse resistivity by sweeping the magnetic field 
between -9~T $\leq$ $\mu_0H$ $\leq$ 9~T applied normal to the film surface, 
and then extract an asymmetric component to determine the Hall 
resistivity $\rho_{xy}$. 
Figures~\ref{Fig.2}(a) - \ref{Fig.2}(c) show the field dependence 
of $\rho_{xy}$ at different temperatures. 
We find that samples A and B show steep increase up to $\mu_0H$ = 2~T, 
which is due to an anomalous Hall effect (AHE). 
$\rho_{xy}$ can be expressed as $\rho_{xy} = B{\cdot}R_H + \rho_s$, 
where $B$ is the magnetic flux density, 
and $\rho_s$ is the anomalous term of the Hall resistivity. 
On the other hand, no trace of AHE is detected in sample C 
evoking that AHE is not a phenomenon intrinsic to FeTe. 
The inset of Fig.~\ref{Fig.2}(d) shows $B$ dependence of $d\rho_{xy}/dB$. 
This gives a rough measure to determine which field range is appropriate 
for the linear fitting to $\rho_{xy}$, 
and we determine to fit the data at 2 $\leq$ $\mu_0H$ $\leq$ 4~T. 
It should be noted that the degree of linearity in high-field region 
is not significantly different among the three films, which indicates 
that the effect of AHE is limited in the weak-field region.

The field-dependence of $\rho_{xy}$ at low temperatures shows contrasting 
results from Fe(Se$_{0.5}$Te$_{0.5}$). 
In our previous study of Fe(Se$_{0.5}$Te$_{0.5}$) thin films,
\cite{Tsukada1} 
$\rho_{xy}$ exhibits a strong nonlinearity to the field at low temperatures, 
which was considered as a strong evidence of the collapse of the condition 
of compensated metals, {\it i.e.} electron density $\neq$ hole density. 
Such behavior is hardly observed in FeTe except for the data 
at limited temperature regions; 
one is around $T$ = 40 - 70~K where the slope of $\rho_{xy}$ changes 
from positive to negative, 
and the other is $T$ = 10~K for sample A only. 
There are two possible reasons for this strong suppression of nonlinearity. 
One is that electron density is equal to hole density, 
and the other is that the mobility of one-type carrier is far larger than 
that of the other carrier. We will discuss it later.

Temperature dependence of $R_H$ is summarized in Fig.~\ref{Fig.2}(d). 
It is surprising that all three films follow almost the same lines 
in the high-temperature region, which indicates that the mixture of 
the (101)-oriented domain does not give a strong influence to $R_H$. 
The inset of Fig.~\ref{Fig.2}(d) demonstrates that at $T$ = 300~K 
the slope of $\rho_{xy}$ is similar to each other 
in spite of the difference of $\rho_s$. 
The deviation from the high-temperature trend becomes obvious below 100~K. 
$R_H$'s start to decrease rapidly, change sign from positive to negative, 
and then are roughly saturated below 30~K. 
The value at the lowest temperature $T$ = 10~K is 
-6.72 $\times$ 10$^{-10}$~m$^3$/C, -2.38 $\times$ 10$^{-9}$~m$^3$/C, 
and -2.71 $\times$ 10$^{-9}$~m$^3$/C for samples A, B, and C, respectively. 
The values for samples B and C are roughly consistent with the recently 
reported numbers in bulk single crystals.
\cite{Liu1}

In order to obtain further insight to the behavior of electrons and holes, 
it is useful to evaluate the mobility of carriers. 
However, it is unrealistic to directly deal with the five conduction bands 
of FeTe. 
We thus adopted a reduced two-band model containing one electron band 
and one hole band. 
In the two-band Drude model, resistivity and Hall coefficients can be 
described as 
$\rho$ = $1 / e(\mu_hn_h + \mu_en_e)$ 
and $R_H$ = $(\mu_h^2n_h - \mu_e^2n_e) / e(\mu_hn_h + \mu_en_e)^2$, 
where $\mu_h$, $\mu_e$, $n_h$, and $n_e$ are hole mobility, electron mobility, 
hole density, and electron density, respectively. 
Once $n_h$ and $n_e$ are given, we can calculate $\mu_h$ and $\mu_e$. 
As was first predicted theoretically,
\cite{Ma1} 
and was later confirmed by neutron-diffraction studies,
\cite{Bao1,Li1} 
FeTe has a bicollinear AFM order in low-temperature region. 
Therefore, we may apply the calculated carrier densities by Ma {\it et al.,} 
\cite{Ma1} 
to our mobility estimation, 
in which $n_h$ ($n_e$) = 2.26 (2.38) $\times$ 10$^{27}$~m$^{-3}$ are reported. 
Ma {\it et al.} also reported the carrier densities in the nonmagnetic state 
as $n_h$ = $n_e$ = 4.77 $\times$ 10$^{27}$~m$^{-3}$ 
(corresponding to a compensated metal),
\cite{Ma1} 
and at first glance these values correspond to the high-temperature state. 
However, the presence of free localized spins above $T_{tm}$ has been reported 
by magnetic susceptibility.
\cite{Xia1,Michioka1} 
This means that the high-temperature region should be regarded 
as a Curie-Weiss type paramagnetic state, and hence those calculated 
for the nonmagnetic state are likely overestimated. 
We thus need to apply the carrier densities calculated for the paramagnetic 
state, 
and use the following values 
$n_h$ ($n_e$) = 1.07 (1.26) $\times$ 10$^{27}$~m$^{-3}$.
which will appear elsewhere in near future.
\cite{Lu1}

%(Mobility1)

\begin{figure}
\begin{center}
\includegraphics*[width=120mm]{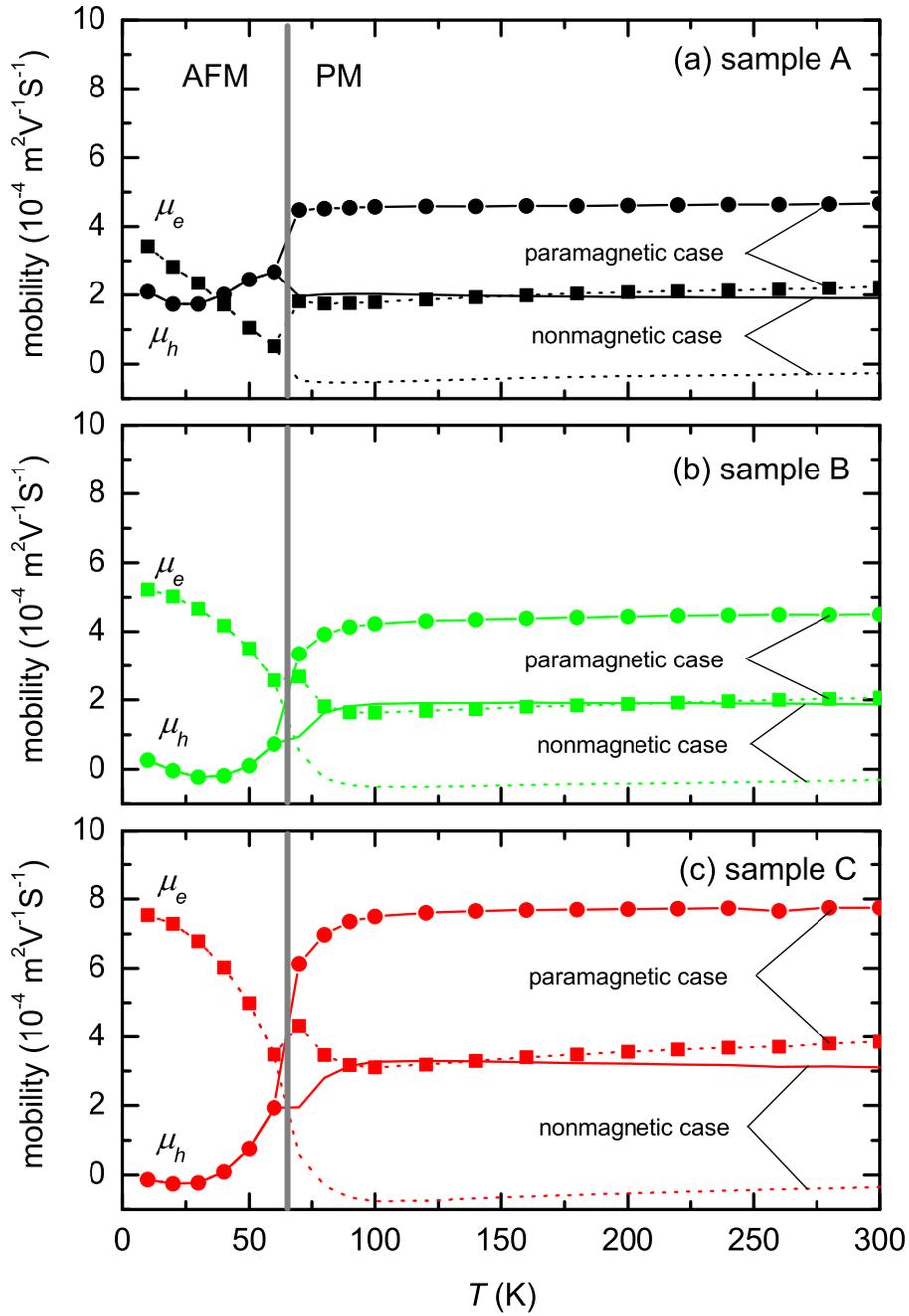}
\end{center}
\caption{(Color online) 
(a)-(c) Mobility of holes and electrons in three cases; 
$n_h$ ($n_e$) = 2.26 (2.38) $\times$ 10$^{27}$~m$^{-3}$ in the bicollinear 
AFM phase at $T$ $\leq$ 60~K, while at $T$ $\geq$ 70~K $n_h$ ($n_e$) = 
1.07 (1.26) $\times$ 10$^{27}$~m$^{-3}$ for the paramagnetic case, 
and $n_h$ = $n_e$ = 4.77 $\times$ 10$^{27}$~m$^{-3}$ in nonmagnetic case 
are assumed. 
}
\label{Fig.3}
\end{figure}

Calculated mobility is summarized in Figs.~\ref{Fig.3}(a)-(c). 
At high-temperature ($T$ $\geq$ 70~K), we plot $\mu_h$ and $\mu_e$ not only 
for the paramagnetic case but also the nonmagnetic case for comparison. 
It is easily seen that the $\mu_e$ takes negative values in the whole 
temperature region in common to all the films, and thus this assumption 
looks unrealistic. 
In the paramagnetic case, we obtained rather reasonable values of mobilities, 
and we discuss them hereafter. 
The electric transport is dominated by holes at high temperatures; 
$\mu_h$ exceeds $\approx$ 4 $\times$ 10$^{-4}$~m$^2$V$^{-1}$s$^{-1}$ 
for samples A and B, while $\mu_h$ of sample C reaches almost 
8 $\times$ 10$^{-4}$~m$^2$V$^{-1}$s$^{-1}$, 
which is the main reason of the lower resistivity of sample C. 
$\mu_e$'s commonly show small values for three samples; $\mu_e$'s 
are approximately half of $\mu_h$, indicating that the contribution 
of $n$-type carrier is strongly suppressed in the paramagnetic (PM) state. 
This is surprising because the band calculation predict almost similar 
band mass for both the electron pocket around $M$ point and the hole pocket 
around $\Gamma$ point.
\cite{Subedi1,Ma1} 
We may infer that this large difference in the character of 
$p$- and $n$-type carriers is dependent on the more detailed structure of 
the Fermi surfaces, which will be clarified with ARPES and/or another methods.

The dominancy of holes is completely interchanged when the antiferromagnetic 
long-range order is evolved. 
Once FeTe goes into the AFM state, $\mu_h$ starts decreasing and 
simultaneously $\mu_e$ starts increasing. 
Strikingly, $\mu_h$ is completely suppressed to zero in samples B and C, 
which indicates that the $p$ type carrier has been almost localized. 
On the other hand, the evolution of $\mu_e$ is remarkable, and its 
magnitude exceeds the value of $\mu_h$ in the PM state. 
This large enhancement of $\mu_e$ is the main reason of metallic, 
($d{\rho} / dT$ $>$ 0), conduction in the AFM state. 
Another important finding is that the suppression of $\mu_h$ is less 
emphasized in sample A. 
The suppression of $\mu_h$ actually occurs in sample A, 
but the minimum value of $\mu_h$ remains at sufficiently high value as 
1.8 $\times$ 10$^{-4}$ m$^2$V$^{-1}$s$^{-1}$, which is comparable to $\mu_e$. 
Therefore, we may expect in sample A that both electrons and holes coexist 
as itinerant carriers even in the AFM state.

%(Discussion1)

The mobility analysis indicates that $\mu_h$ remains finite in superconducting 
sample while it goes to almost zero in nonsuperconducting samples. 
In other words, sufficient itinerancy is necessary for the occurrence 
of superconductivity in both electron and hole bands. 
This result prefers a scenario of pairing mechanism requiring 
an interband scattering between $p$- and $n$-type Fermi surfaces, 
and hence supports either $s_{++}$-wave or $s_{\pm}$-wave pairing states, 
\cite{Mazin1,Kuroki1,Kontani1} 
while it may exclude the possibility of simple anisotropic $s$- and $d$-wave 
pairing state in which gap opens independently on each Fermi surface. 
One may claim that the strong suppression of $\mu_h$ to zero is 
just an artifact of the calculations. 
However, we have confirmed that a slight change in the carrier density 
never alters the tendency that superconducting sample exhibits higher 
hole mobility than that nonsuperconducting one does, 
and hence, the correlation of superconductivity with the itinerancy 
of holes and electrons seems to be robust. 
It should be noted that our analysis taking only two bands into account 
does not loose generality for the real five band transport. 
We should emphasize again that the present results discuss an extreme case 
where $\mu_h$ is almost zero at low temperatures. 
In this case, we may safely conclude that all holes are not mobile, 
and the qualitative discussion carried out before is still valid.

%(summary)

In conclusion, we measured the Hall effect for nonsuperconducting and 
superconducting FeTe thin films in detail. 
Temperature dependence of the Hall coefficients is analyzed by a semiclassical 
two-band Drude model with the aid of calculated carrier density. 
The estimated mobility of $n$- and $p$-type carriers 
demonstrates a remarkable interchange across the antiferromagnetic transition. 
In the antiferromagnetic state, the mobility of $p$-type carriers of 
nonsuperconducting FeTe is suppressed to almost zero, 
while it remains finite in superconducting FeTe, 
which indicates that the necessity condition for the occurrence 
of superconductivity is that both electrons and holes remains itinerant.

%(Acknowledgment)
The authors are grateful to Z. -Y. Lu and Tao Xiang for sharing with us 
the unpublished data prior to publication. 
We also thank H. Kontani, T. Tohyama, and  K. Ohgushi fruitful discussions.

%Long Reference

\newpage

\newpage

\newpage

\end{document}